\documentclass[amsfonts,prd,aps]{revtex4}

\usepackage{graphicx}

\begin{document}

\title{Emergent diffeomorphism invariance in a discrete loop
quantum gravity model}

\author{ Rodolfo Gambini$^{1}$,
Jorge Pullin$^{2}$}
\affiliation {
1. Instituto de F\'{\i}sica, Facultad de Ciencias, 
Igu\'a 4225, esq. Mataojo, Montevideo, Uruguay. \\
2. Department of Physics and Astronomy, Louisiana State University,
Baton Rouge, LA 70803-4001}

\begin{abstract}
  Several approaches to the dynamics of loop quantum gravity involve
  discretizing the equations of motion. The resulting discrete
  theories are known to be problematic since the first class algebra
  of constraints of the continuum theory becomes second class upon
  discretization.  If one treats the second class constraints
  properly, the resulting theories have very different dynamics and
  number of degrees of freedom than those of the continuum theory.  It
  is therefore questionable how these theories could be considered a
  starting point for quantization and the definition of a continuum
  theory through a continuum limit. We show explicitly in a model that
  the {\em uniform discretizations} approach to the quantization of
  constrained systems overcomes these difficulties.  We consider here
  a simple diffeomorphism invariant one dimensional model and complete
  the quantization using {\em uniform discretizations}. The model can
  be viewed as a spherically symmetric reduction of the well known
  Husain--Kucha\v{r} model of diffeomorphism invariant theory. We show
  that the correct quantum continuum limit can be satisfactorily
  constructed for this model. This opens the possibility of treating
  $1+1$ dimensional dynamical situations of great interest in quantum
  gravity taking into account the full dynamics of the theory and
  preserving the space-time covariance at a quantum level.
\end{abstract}

\maketitle
\section{Introduction}

Lattice techniques have proved remarkably useful in the quantization of 
usual gauge theories.
This raised the hope that they may also prove useful in the quantization
of gravity. A major difference however is that most theories of gravity of
interest are invariant under diffeomorphisms and the introduction of a
discrete structure breaks diffeomorphism invariance. One of the appealing
features of lattice gauge theories is therefore lost in this case, one
breaks the symmetry of the theory of interest. The situation gets further
compounded in the case of canonical general relativity, since there one
also breaks four dimensional covariance into a $3+1$ dimensional split.
Spatial diffeomorphisms get implemented via a constraint that has a 
natural geometrical action and the usual algebra of diffeomorphisms
is implemented via the constraint algebra. But the remaining space-time
diffeomorphism gets implemented through the complicated Hamiltonian
constraint, that has a challenging algebra with spatial diffeomorphisms.
In particular the algebra of constraints has structure functions. 
If we call $C(\vec{N})$ the diffeomorphism constraint smeared by
a test vector field (shift) $\vec{N}$ and $H(N)$ the Hamiltonian constraint
smeared by a scalar lapse $N$, the constraint algebra is,
\begin{eqnarray}
\left\{C(\vec{N}),C(\vec{M})\right\}=C([\vec{N},\vec{M}])\\
\left\{C(\vec{N}),H(M)\right\}=H({\cal L}_{\vec{N}}M)\\
\left\{H({N}),H(M)\right\}=C(\vec{K}(q)),
\end{eqnarray}
where the vector $K^a=q^{ab}(N\partial_a M-M\partial_a N)$
and $q^{ab}$ is the spatial metric. The last Poisson bracket therefore
involves structure functions depending on the canonical variables
on the right hand side.

The algebra of constraints poses important complications in the context of
loop quantum gravity when one wishes to implement it as an operator
algebra at a quantum level (see \cite{ThiemannGiesel} for a lengthier
discussion). In particular, if one chooses spin network
states with the usual Ashtekar-Lewandowski \cite{AsLe} measure, they form a
non-separable Hilbert space. In it, diffeomorphisms are not
implemented in a weakly continuous fashion, i.e. finite
diffeomorphisms can be represented but infinitesimal ones cannot. This
implies that in loop quantum gravity one treats very asymetrically the
spatial and temporal diffeomorphisms. Whereas invariance under spatial
diffeomorphisms is implemented via a group averaging procedure 
\cite{groupaveraging},
invariance under the remaining space-time diffeomorphisms is to be
implemented by solving a quantum operatorial equation corresponding to
the Hamiltonian constraint.  Since the Poisson bracket of two
Hamiltonian constraints involves the infinitesimal generator of
diffeomorphisms, which is not well defined as a quantum operator, one
cannot expect to implement the Poisson algebra at an operatorial level
in the quantum theory, at least in the kinematical Hilbert space. 

A symmetric treatment of the diffeomorphism and Hamiltonian constraints
requires to develop a technique that allows to implement the generators
of spatial diffeomorphisms as operators in the loop representation. 
One could attempt to treat the diffeomorphism and Hamiltonian 
constraints on the same footing, for instance by lattice regularizing
them. Unfortunately, such discretized versions of the constraints 
are not first class. If one treats them properly with the Dirac
procedure, the resulting theory is vastly different in symmetries
and even in the number of degrees of freedom from what one expects
to have in the continuum theory. Therefore there is little chance
that one could define a continuum theory as a suitable limit of the
constructed lattice theories.

These problems have led to the consideration of extensions of the
Dirac procedure that could better accommodate this particular problem
with the constraint algebra. One such approach is the ``master 
constraint'' programme of Thiemann and collaborators \cite{master}. Another 
approach that we have been studying in the last few years are the
``uniform discretizations'' \cite{uniform}. Both approaches have 
some elements in common.

Uniform discretizations are discrete versions of a constrained theory
in which the discretized form of the constraints are quantities whose
values are under control throughout the system's evolution.  Notice
that this would not be the case, for instance, if one simply takes a
constrained theory and discretizes it. Initial data on which the
discrete version of the constraints vanishes will evolve into data
with non-vanishing values of the discrete constraints, without any
control on the final value. This situation is well known, for
instance, in numerical relativity.  Uniform discretizations are
designed in such a way that the discrete constraints are kept under
control upon evolution and that one can take appropriate limits in the
initial data such that one can satisfy the constraints to an arbitrary
(and controlled) degree of accuracy. This therefore guarantees the
existence of a well defined continuum limit at the level of the
classical theory. It has been shown \cite{discreteexamples} that the
uniform discretization technique is classically equivalent to the
Dirac procedure when the constraints are first class. For second class
constraints, like the ones that arise when one discretizes continuum
systems with first class constraints the uniform discretization
technique is radically different from the Dirac procedure, yielding a
dynamical evolution that recovers in the continuum limit the continuum
theory one started with.

Although the existence of a continuum limit is generically guaranteed at a
classical level, it is not obvious that it is at the quantum level. It is
known \cite{discreteexamples} that there are models in which the
continuum limit cannot be achieved and one is left with a non-zero
minimum value of the expectation value of the sum squared of the
constraints. It is therefore of interest to show that in examples of
growing complexity and of increasing similarity to general relativity
one can indeed define a continuum quantum theory with the desired
symmetries by applying the uniform discretization procedure. The purpose
of this paper is to discuss one such model. We will consider the
quantization via uniform discretizations of a $1+1$ dimensional model
with diffeomorphism symmetry and we will show that the symmetry is
recovered at the quantum level correctly. This raises the hopes of having
a theory where all the constraints are treated on an equal footing.

The organization of this paper is as follows. In section II we discuss
the model we will consider. In section III we 
discretize the model. In section IV we review the
uniform discretization procedure and how it departs from the 
Dirac traditional approach. Section VI discusses the quantization
using uniform discretizations and how one recovers the correct
continuum limit. We conclude with a discussion.

\section{The model}

We would like to construct a model by considering spherically
symmetric gravity and ignoring the Hamiltonian constraint.  This is
analogous to building a ``Husain--Kuchar'' \cite{husainkuchar} version
of spherically symmetric gravity. It is known that these models
correspond to degenerate space-times when translated in terms of the
metric variables.

We refer the reader to our previous work on spherically symmetric
gravity \cite{spherical} 
for the setup of the model in terms of Ashtekar's new variables.
Just as a recap, the model has two canonical pairs $K_x, E^x$ and
$K_\varphi,E^\varphi$. The relation to the more traditional 
metric canonical variables is,
\begin{eqnarray}
g_{xx}&=& \frac{(E^\varphi)^2}{|E^x|},\qquad g_{\theta\theta} = |E^x|,\\
K_{xx}&=&-{\rm sign}(E^x) \frac{(E^\varphi)^2}{\sqrt{|E^x|}}K_x,\qquad
K_{\theta\theta} = -\sqrt{|E^x|} {K_\varphi}
\end{eqnarray}
and we have set the Immirzi parameter to one for simplicity, since it
does not play a role in this analysis.

The Lagrangian for spherically symmetric gravity ignoring the Hamiltonian
constraint is,
\begin{equation}
L = \int dx E^x \dot{K}_x+E^\varphi \dot{K}_\varphi +N 
\left((E^x)'K_x - E^\varphi (K_\varphi)'\right)
\end{equation}
with $N$ a Lagrange multiplier (the radial component of the shift
vector). The equations of motion are
\begin{eqnarray}
\dot{K}_x-\left(NK_x\right)' &=&0,\\
\dot{E}_x-N\left(E^x\right)' &=&0,\\
\dot{K}_\varphi-NK'_\varphi &=&0,\\
\dot{E}^\varphi-\left(NE^\varphi\right)' &=&0.
\end{eqnarray}
The theory has one constraint, which is the remaining diffeomorphism
constraint in the radial $(x)$ direction, $\phi=
-\left(E^x\right)'K_x+E^\varphi K'_\varphi$, which we will write smeared
as $\phi(N)=\int dx N \phi$.  The constraint generates diffeomorphisms
of the fields, with $K_\varphi$ and $E^x$ behaving as scalars and $K_x$ and $E^\varphi$  as a densities of weight one,
\begin{eqnarray}
\delta K_\varphi &=& \left\{K_\varphi,\phi(N)\right\}=N K'_\varphi,\\
\delta K_x &=& \left\{K_x,\phi(N)\right\}=\left(N K_x\right)',\\
\delta E^\varphi &=& \left\{E^\varphi\phi(N)\right\}=\left(N E^\varphi\right)',\\
\delta E^x &=& \left\{E^x,\phi(N)\right\}=N \left(E^x\right)'.
\end{eqnarray}
The constraint has the usual algebra of diffeomorphisms,
\begin{equation}
\left\{\phi(N),\phi(M)\right\}=\phi\left(N M'-M N'\right).
\end{equation}

Observables are integrals of densities of weight one constructed
with the fields, for example, $O=\int dx f(E^x,K_\varphi)K_x$ with $f$ a function.
One then has
\begin{equation}
\left\{O, \phi(N)\right\}=\int dx \left[\frac{\partial f}{\partial E^x} N \left(E^x\right)'
+\frac{\partial f}{\partial K_\varphi} N K'_\varphi+ \left(NK_x\right)' f\right] =
\int dx \partial_x \left(f NK_x\right)=0,
\end{equation}
if one considers a compact spatial manifold, $S^{1}$, which we will do
throughout this paper. (This may not make a lot of sense if one is thinking
of the model as a reduction of $3+1$ spherical symmetry, but we are just
avoiding including  boundary terms, which are straightforward to
treat in the spherical case, see \cite{spherical}, in order to simplify the
discussion of diffeomorphism invariance).

\section{Discretization}

We now proceed to discretize the model. The spatial direction $x$ is 
discretized into points $x_i$ such that $x_{i+1}-x_i=\epsilon_i$ 
and the distances are smaller than a bound $d(\epsilon_i)<d_\epsilon$ when 
measured in  some fiducial metric. To simplify notation, from now on
we will assume the points are equally spaced and drop the suffix $i$ on
$\epsilon$, but the analysis can be straightforwardly extended to the case
with variable $\epsilon_i$.
The variables of the model 
become $K_{x,i}=K_x(x_i)$, $K_{\varphi,i}=K_\varphi(x_i)$ 
and $E^x_i=\epsilon E^x(x_i)$ and
$E^\varphi_i=\epsilon E^\varphi(x_i)$. The constraint is,
\begin{equation}
\phi_i =E^\varphi_i\left(K_{\varphi,i+1}-K_{\varphi,i}\right) -K_{x,i}
\left(E^x_{i+1}-E^x_i\right).
\end{equation}
The constraint algebra is not first class, i.e.,
\begin{eqnarray}
\left\{\phi_i,\phi_j\right\}&=&-E^\varphi_{i-1}\left(K_{\varphi,i+1}-K_{\varphi,i}
\right)
\delta_{i,j+1}+E^\varphi_{j-1}\left(K_{\varphi,j+1}-K_{\varphi,j}
\right)\delta_{j,i+1}\nonumber\\
&&K_{x,{i-1}}\left(E^{x}_{i+1}-E^x_i
\right)
\delta_{i,j+1}-K_{x,j-1}\left(E^x_{j+1}-E^x_j
\right)\delta_{j,i+1} 
\end{eqnarray}
which does not reproduce the constraint. What one has is a ``classical
anomaly'' of the form $\left(E^\varphi_{i+1}-E^\varphi_{i}\right)
\left(K_{\varphi,i}-K_{\varphi,i-1}\right)
-\left(E^x_{i+1}-E^x_i\right)\left(K_{x,i}-K_{x,i-1}\right)$. 
These terms would tend to zero if one takes the separation $\epsilon$
to zero and the variables behave continuously in such a limit.

So if one were to simply quantize the discrete model, one would run
into trouble since one would be quantizing a classical theory with
second class constraints. We will expand more on the problems one faces in the
next section. In this paper we would
like to show that in spite of this problem of the classical theory, which
implies that the discrete theory loses diffeomorphism invariance, if one
follows the uniform discretization approach to quantization the 
diffeomorphism invariance is recovered in the limit $\epsilon\to 0$ both
at the classical and quantum level.

In the uniform discretization approach one constructs a ``master
constraint'' ${\mathbb H}$ by considering the sum of the discretized
constraints squared.  One then promotes the resulting quantity to a
quantum operator and seeks for the eigenstates of $\hat{\mathbb H}$
with minimum eigenvalue. In the full theory the quantity ${\mathbb H}$
would be constructed from the diffeomorphism constraints $\phi_a$ as,
\begin{equation}
{\mathbb H} =\frac{1}{2} \int dx \phi_a \phi_b \frac{g^{ab}}{\sqrt{g}},
\end{equation}
which motivates in our example to choose,
\begin{equation}
{\mathbb H} = \frac{1}{2}\int dx \phi \phi \frac{\sqrt{E^x}}{\left(E^\varphi\right)^3},
\end{equation}
or, in the discretized theory as,
\begin{equation}
{\mathbb H}^\epsilon = \frac{1}{2}\sum_{i=0}^N \phi_i \phi_i
\frac{\sqrt{E^x_i}}{\left(E^\varphi_i\right)^3} \epsilon^{3/2}.
\end{equation}

To understand better how to promote these quantities to quantum operators,
it is best to start with the constraint itself. Let us go back for a second
to the continuum notation, and write,
\begin{equation}
\phi^\epsilon(N)= \sum_{j=0}^N \epsilon N(x_j) \left\{
-\frac{\left[E^x(x_{j+1})-E^x(x_j)\right]}{\epsilon} K_x(x_j)+
\frac{1}{2}\left[E^\varphi(x_{j})+E^\varphi(x_{j+1})\right]
\frac{\left(K_\varphi(x_{j+1})-K_\varphi(x_j)\right)}{\epsilon}\right\} ,
\end{equation}
which would reproduce the constraint $\phi(N)=\lim_{\epsilon\to0}
\phi^\epsilon(N)$ though  we see that the explicit 
dependence on $\epsilon$ drops out. We have chosen to regularize
$E^\varphi$ at the midpoint in order to simplify the action of the 
resulting quantum operator as we will see later. When one is to promote these
quantities to quantum operators, one needs to remember that although
the $E$ variables promote readily to quantum operators in the loop
representation, the $K$'s need to be written in exponentiated form. To
this aim, we write, classically,
\begin{equation}
\phi^\epsilon(N)= \sum_{j=0}^N \frac{N(x_j)}{2i\epsilon}\left\{
\exp\left(-2i\epsilon{\left[E^x(x_{j+1})-E^x(x_j)\right]} K_x(x_j)+
i\epsilon{\left[E^\varphi(x_{j})+E^\varphi(x_{j+1})\right]}
\left(K_\varphi(x_{j+1})-K_\varphi(x_j)\right)\right)-1\right\},
\end{equation}
which again would reproduce the constraint in the continuum limit.
Let us rewrite it in terms of the discrete variables,
\begin{equation}
\phi^\epsilon(N)= \sum_{j=0}^N \frac{N_j}{2i\epsilon}\left\{
\exp\left[i\left({-2\left[E^x_{j+1}-E^x_j\right]} K_{x,j}+
{\left[E^\varphi_{j}+E^\varphi_{j+1}\right]}
\left(K_{\varphi,j+1}-K_{\varphi,j}\right)\right)\right]-1\right\}.
\end{equation}

For later use, it is convenient to rewrite $\phi_j^\epsilon 
= (D_j-1)/(2i\epsilon)$ and then 
one has that,
\begin{equation}
{\mathbb H}^\epsilon = \sum_{j=0}^N
 \left(D_j-1\right)\left(D_j-1\right)^* \epsilon^{-1/2} \frac{\sqrt{E^x_j}}
{\left(E^\varphi_j\right)^3}.\label{27}
\end{equation}
We dropped the $\epsilon$ in $D$ since it does not explicitly depend
on it, but it does through the dependence on $E^x$ and an irrelevant
global factor of $1/8$ to simplify future expressions.

\section{Uniform discretizations}

Before quantizing, we will study the classical theory using 
uniform discretizations and we will verify that one gets
in the continuum limit a theory with diffeomorphism constraints
that are first class. The continuum theory can be treated with
the Dirac technique and has first class constraints that 
generate diffeomorphisms on the dynamical variables. However,
the discrete theory, when treated with the Dirac technique, 
has second class constraints and does not have the gauge
invariances of the continuum theory. The number of degrees
of freedom changes and the continuum limit generically
does not recover the theory one started with.

As mentioned before, it has been shown \cite{discreteexamples} 
that the uniform discretization technique is
equivalent to the Dirac procedure when the constraints are first 
class. For second class constraints, like the ones that appear when
one discretizes continuum systems with first class constraints the
uniform discretization technique is radically different from the
Dirac procedure, yielding a dynamical evolution that recovers in 
the continuum limit the continuum theory one started with. 

Let us review how this works. We start with a classical canonical
system with $N$ configuration variables, parameterized by a continuous
parameter $\alpha$ such that $\alpha\to 0$ is the ``continuum limit''.
We will assume the theory in the continuum has $M$ constraints $\phi_j
= \lim_{\alpha\to 0} \phi_j^\alpha$.  In the discrete theory we will
assume the constraints generically fail to be first class,
\begin{equation}
  \left\{\phi^\alpha_j,\phi^\alpha_k\right\} =
\sum_{m=1}^M C^\alpha_{jk}{}^m \phi^\alpha_m+ A^\alpha_{jk},
\end{equation}
where the failure is quantified by $A^\alpha_{jk}$. We assume
that in the continuum limit one has $\lim_{\alpha\to 0} A^\alpha_{jk}=0$
and that the quantities $C^\alpha_{jk}{}^m$ become in the limit the
structure functions of the (first class) 
constraint algebra of the continuum theory $C_{jk}{}^m=\lim_{\alpha\to 0}
C^\alpha_{jk}{}^m$, so that,
\begin{equation}
\left\{\phi_j,\phi_k\right\} =\sum_{m=1}^M C_{jk}{}^m \phi_m.
\end{equation}

If one were to insist on treating the above discrete theory using the
Dirac procedure, that is, taking the constraints $\phi^\alpha_j=0$ and
a total Hamiltonian $H_{T}=\sum_{j=1}^M C_j \phi^\alpha_j$ with $C_j$
functions of the canonical variables, one immediately finds
restrictions on the $C_j{}'s$ of the form $\sum_{j=1}^M C_j A^\alpha_{jk}=0$ in order to
preserve the constraints upon evolution. Only in the continuum
$\alpha\to 0$ limit are the $C_j$ free functions and one has in the
theory $2N-2M$ observables. Notice that away from  the continuum limit
the number of observables is generically larger and could even reach
$2N$ if the matrix $A^\alpha_{jk}$ is invertible. Therefore one cannot
view the theory in the $\alpha\to 0$ limit as a limit of the theories
for finite values of $\alpha$, since they do not even have the same
number of observables and have a completely different evolution.

The uniform discretizations, on the other hand, lead to discrete
theories that have the same number of observables and an
evolution resembling those of the continuum theory. One can then
claim that the discrete theories approximate the continuum theory
and the latter arises as the continuum limit of them.

The treatment of the system in questions would start with 
the construction of the ``master constraint''
\begin{equation}
{\mathbb H}^\alpha=\frac{1}{2} \sum_{i=j}^M \left(\phi^\alpha_j\right)^2
\end{equation}
and defining a discrete time evolution through ${\mathbb H}$. In 
particular, this implies a discrete time evolution from instant $n$ to $n+1$ 
for the 
constraints of the form,
\begin{eqnarray}
  \phi^\alpha_j(n+1) &=& \phi^\alpha_j(n)+ \label{31}
\left\{ \phi^\alpha_j(n),{\mathbb H}^\alpha\right\}+
\frac{1}{2} 
\left\{\left\{ \phi^\alpha_j(n),{\mathbb H}^\alpha\right\},
{\mathbb H}^\alpha\right\}+
\ldots\\
&=&\phi^\alpha_j(n)+ \sum_{i,k=1}^M C^\alpha_{ji}{}^k \phi^\alpha_k(n)
\phi^\alpha_i(n)+
\sum_{i=1}^M A^\alpha_{ji} \phi^\alpha_i(n)+\ldots\label{evolution}
\end{eqnarray}
This evolution implies that ${\mathbb H}^\alpha$ is a constant of the motion,
which for convenience we denote via a parameter $\delta$ such that
${\mathbb H}^\alpha=\delta^2/2$. The preservation upon evolution of 
${\mathbb H}^\alpha$ implies that the constraints 
remain bounded $|\phi^\alpha_j|\leq
\delta$. 

If one now divides  by $\delta$ and defines the quantities
$\lambda^\alpha_i\equiv 
\phi^\alpha_i/\delta$ one can rewrite (\ref{evolution}) as,
\begin{equation}
\frac{\phi^\alpha_j(n+1)-\phi^\alpha_j(n)}{\delta}= 
\sum_{i,j=1}^M C^\alpha_{ji}{}^k \phi^\alpha_k(n)
\lambda^\alpha_i(n)+
\sum_{i,j=1}^M A^\alpha_{ji} \lambda^\alpha_i(n)+\ldots
\end{equation}
Notice that the $\lambda^\alpha_i$ remain finite when one takes the limits
$\delta\to 0$ and $\alpha\to 0$.

If one now considers the limit of small $\delta$'s, one notes that the
first term on the right is of order $\delta$, the second one goes to zero
with  $\alpha\to 0$, at least as $\alpha$ and the rest of the terms are
of higher orders in $\delta,\alpha$. If one identifies with a continuum
variable $\tau$ such that $\tau=n\delta+\tau_0$, then
$\phi^\alpha_j(\tau)\equiv\phi^\alpha_j(n)$ and $\phi^\alpha_j(\tau+\delta)\equiv\phi^\alpha_j(n+1)$ one can
take the limits $\alpha\to 0$ and 
$\delta\to 0$, irrespective
of the order of the limits one gets that the evolution equations (\ref{evolution}) for the constraints become those of the continuum theory, i.e.,
\begin{equation}
\dot{\phi}_j \equiv \lim_{\alpha,\delta\to 0} 
\frac{\phi^\alpha_j(\tau+\delta)-\phi^\alpha_j(\tau)}{\delta}= 
\sum_{i,k=1}^M C_{ji}{}^k \phi_k \lambda_i
\end{equation}
with $\lambda_i$ become the (freely specifiable) Lagrange multipliers of
the continuum theory. At this point the reader may be puzzled, since the 
$\lambda$'s are defined as limits of those of the discrete theory and therefore
do not appear to be free. However, one has to recall that the $\lambda$'s in
the discrete theory are determined by the values of the constraints evaluated
on the initial data, and these can be chosen arbitrarily by modifying the
initial data.

If one considers the limit $\delta\to 0$ for a finite value of $\alpha$
(``continuous in time, discrete in space'') and considers the evolution
of a function of phase space $O$, one has that,
\begin{equation}
  \dot{O}=\left\{O,{\mathbb H}^\alpha\right\}=\left\{O,\phi^\alpha_i\right\}\lambda^\alpha_i+\sum_{j=1}^M
\left\{O,\phi^\alpha_i\right\}A^\alpha_{ij} \lambda^\alpha_j 
+\sum_{j,k=1}^M\left\{O,\phi^\alpha_i\right\} A^\alpha_{ij}A^\alpha_{jk} \lambda^\alpha_k+\ldots
\end{equation}

The necessary and sufficient condition for $O$ to be a constant of the motion (that is, $\dot{O}=0$) is that
\begin{equation}
\left\{O,\phi^\alpha_i\right\}=\sum_{j=1}^M C_{ij} \phi^\alpha_j+B^\alpha_i,
\end{equation}
with $B^\alpha_i$ a vector, perhaps vanishing, that is annihilated by the
matrix,
\begin{equation}
\Lambda^\alpha_{ij} = \delta_{ij} + A^\alpha_{ij}+
\sum_{k=1}^MA^\alpha_{ik}A^\alpha_{kj}+\ldots+\sum_{k_1=1,\ldots,k_s=1}^M A^\alpha_{i,k_1}\cdots
A^\alpha_{k_s,j}+\ldots
\end{equation}

Up to now we have assumed 
$\lambda^\alpha_i$ arbitrary and not necessarily satisfying that  $\sum_{j=1}^N A^\alpha_{ij}
\lambda_j =0$. It is clear that $\lim_{\alpha\to 0} \Lambda^\alpha_{ij}=\delta_{ij}$ and therefore 
$\lim_{\alpha\to 0} B^\alpha_i = 0$ which implies that conserved quantities in the discrete theory
yield in the limit $\alpha\to 0$ the observables of the continuum theory. 

Since the $\lambda_i$'s are free the theory with continuous time is
not the one that would result naively from applying the Dirac
procedure since in the latter the Lagrange multipliers are restricted by 
$\sum_{j=1}^M A_{ij} \lambda^\alpha_j=0$
and therefore the theory  admits more observables than the $2N-2M$ of the
continuum theory. That is, if one takes the ``continuum in time'' limit
first, the discrete theory has a dynamics that differs from the usual
one unless $A^\alpha_{ij} \phi^\alpha_i(n)=0$ and one is really
treating two different theories.

At this point it would be good to clarify a bit the notation. The above
discussion has been for a mechanical system with $M$ configuration 
degrees of freedom. When one discretizes a field theory with $M$
configuration degrees of freedom on a lattice with $N$ points one
ends up with a mechanical system that has $M\times N$ degrees of 
freedom.  An example of such a system would be the
diffeomorphism constraints of general relativity in $3+1$ dimensions
when discretized on a uniform lattice of spacing $\alpha$
\cite{rentelnsmolin}. Of course, it is not clear at this point if such
a system could be completely treated with our technique up to the last
consequences, we just mention it here as an example of the type of
system one would like to treat. The above discussion extends
immediately to systems of this kind, only the bookkeeping has to be
improved a bit. If we consider a parameter $\alpha(N)=1/N$, such that
the continuum limit is achieved in $N\to\infty$ the classical
continuum constraints can be thought of as limits
\begin{equation}
  \phi_j(x)= \lim_{N\to\infty} \phi^{\alpha(N)}_{j,i(x,N)}
\end{equation}
where $i(x,N)$ is such that the point $x$ in the continuum lies
between $i(x,N)$ and $i(x,N)+1$ on the lattice for every $N$. We 
are assuming a one dimensional lattice. Similar bookkeepings can be
set up in higher dimensional cases.

Just like we did in the mechanical system we can define
\begin{equation}
  \left\{ \phi^{\alpha(N)}_{j,i},\phi^{\alpha(N)}_{k,i\pm 1}\right\}
=\sum_{l,m=1}^MC^{\alpha(N)}_{j,i,k,i\pm1}{}^{lm}\phi^{\alpha(N)}_{l,m} +A^{\alpha(N)}_{j,i,k,i\pm1},
\end{equation}
(where we have assumed that for the sites different 
from $i\pm 1$ on the lattice the Poisson
bracket vanishes, the generalization to other cases is immediate) 
and one has that
\begin{equation}
  \lim_{N\to \infty} A^{\alpha(N)}_{j,i,k,i\pm 1}=0.
\end{equation}
If one takes the spatial limit $\alpha\to 0$ first, one has a theory
with discrete time and continuous space and with first class
constraints and we know in that case the uniform discretization
procedure agrees with the Dirac quantization.

If one has more than one spatial dimension to discretize, then the
situation complicates, since the continuum limit can be achieved with
lattices of different topologies and connectivity. Once one has chosen
a given topology and connectivity for the lattice, the continuum limit
will only produce spin networks of connectivities compatible with such
lattices. For instance if one takes a ``square'' lattice in terms of
connectivity in two spatial dimensions, one would produce at most spin
networks in the continuum with four valent vertices. If one takes a
lattice that resembles a honeycomb with triangular plaquettes one
would produce sextuple vertices, etc. It is clear that this point
deserves further study insofar as to how to achieve the continuum
limit in theories with more than one spatial dimension.

In addition to this, following the uniform discretization approach one
does not need to modify the discrete constraint algebra since it
satisfies $\lim_{N\to\infty} \left\{\phi_i,\phi_j\right\}\sim 0$ and
all the observables of the continuum theory arise by taking the
continuum limit of the constants of the motion of the discrete theory.
The encouraging fact that we recover the continuum theory in the limit
classically is what raises hopes that a similar technique will also
work at the quantum level.

\section{Quantization}

To proceed to quantize the model, we need to consider the master
constraint given in equation (\ref{27}),
\begin{equation}
{\mathbb H}^\epsilon = \sum_{j=0}^N
 \left(D_j-1\right)\left(D_j-1\right)^* \epsilon^{-1/2} \frac{\sqrt{E^x_j}}
{\left(E^\varphi_j\right)^3},
\end{equation}
and quantize it. The quantization of this expression will require appropriate 
ordering of the 
exponential that appears in $D_j$ , putting the $K$'s to the left of the $E$'s, as in usual 
normal ordering. One would then have,
\begin{equation}
\hat{D}_j
= 
:\exp i\left({-2\left[\hat{E}^x_{j+1}-\hat{E}^x_j\right]} \hat{K}_{x,j}+
{\left[\hat{E}^\varphi_{j}+\hat{E}^\varphi_{j+1}\right]}
\left(\hat{K}_{\varphi,j+1}-\hat{K}_{\varphi,j}\right)\right):
\label{D}
\end{equation}
Notice that $\hat{D}_j$ is not self-adjoint and, due to the factor
ordering, neither is $\hat{\phi}_j$, but we will see that one can
construct an $\mathbb H$ that is self-adjoint.

To write the explicit action, let us recall the nature of the basis of
spin network states in one dimension (see \cite{spherical} for
details). One has a lattice of points $j=0\ldots N$.  On such lattice
one has a graph $g$ consistent of a collection of links $e$ connecting
the vertices $v$. It is natural to associate the variable $K_x$ with
links in the graph and the variable $K_\varphi$ with vertices of the
graph. For bookkeeping purposes we will associate each link with the
lattice site to its left. One then constructs the ``point holonomies''
for both variables as,
\begin{equation}
T_{g,\vec{k},\vec{\mu}}(K_x,K_\varphi) = \langle K_x,K_\varphi
\left\vert\vphantom{\frac{1}{1}}\right.\left.
\raisebox{-5mm}{\includegraphics[height=1.5cm]{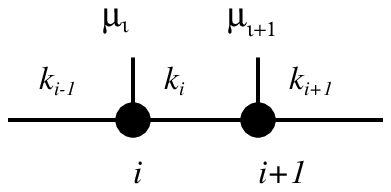}}\right\rangle=
\exp\left(i\sum_{j} {k_j} K_{x,j}\epsilon \right)
\exp\left(i\sum_j \mu_{j,v} K_{\varphi,j}\right)
\end{equation}
The summations go through all the points in the lattice and we allow
the possibility of using ``empty'' links to define the graph, i.e. links where
$k_j=0$.
The vertices of the graph therefore correspond to 
lattice sites where one of the two following conditions are met:
either $\mu_i\neq0$  or $k_{i-1}\neq k_i$. 

In terms of this basis it is straightforward to write the action of the
operator defined in  (\ref{D}),
\begin{eqnarray}
\hat{D}_i 
\left\vert\vphantom{\frac{1}{1}}\right.\left.
\raisebox{-5mm}{\includegraphics[height=1.5cm]{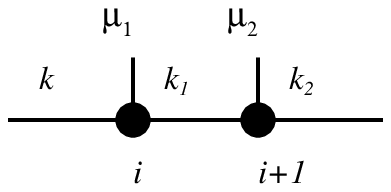}}
\right\rangle
&=&
\left\vert\vphantom{\frac{1}{1}}\right.\left.
\raisebox{-5mm}{\includegraphics[height=1.5cm]{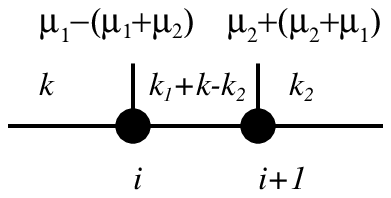}}
\right\rangle
\\
&=&
\left\vert\vphantom{\frac{1}{1}}\right.\left.
\raisebox{-5mm}{\includegraphics[height=1.5cm]{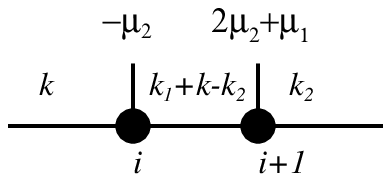}}
\right\rangle.
\end{eqnarray}
The above expression is easy to obtain, since the $\hat{E}^\varphi_j$ may
be substituted by the corresponding eigenvalues $\mu_j$ and 
$\hat{E}^x_j$ produces $(k_{j-1}+k_j)/(2\epsilon)$. 
The exponential of $\lambda K_{\varphi,j}$ 
adds $\lambda$ to $\mu_j$, whereas the exponential of 
$\epsilon n K_{x,i}$ adds $n$ to $k_i$.

An interesting particular case is that of an isolated $\mu$ populated
vertex,
\begin{equation}
\hat{D}_i 
\left\vert\vphantom{\frac{1}{1}}\right.\left.
\raisebox{-5mm}{\includegraphics[height=1.5cm]{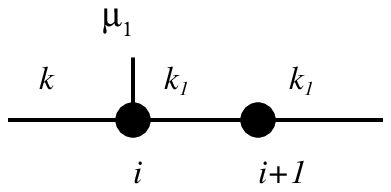}}
\right\rangle
=
\left\vert\vphantom{\frac{1}{1}}\right.\left.
\raisebox{-5mm}{\includegraphics[height=1.5cm]{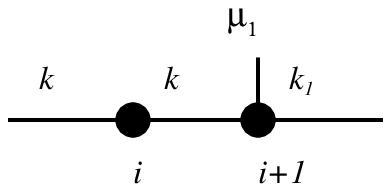}}
\right\rangle.
\end{equation}
So we see that the operator $\hat{D}$ moves the line to a new
vertex. This clean action is in part due to the choice of ``midpoint''
regularization we chose for the $E^\varphi$. This  will in the end
be important to recover diffeomorphism invariance in the continuum.

Something we will have to study later is the possibility of ``coalescing''
two vertices, as in the case,
\begin{equation}
\hat{D}_i 
\left\vert\vphantom{\frac{1}{1}}\right.\left.
\raisebox{-5mm}{\includegraphics[height=1.5cm]{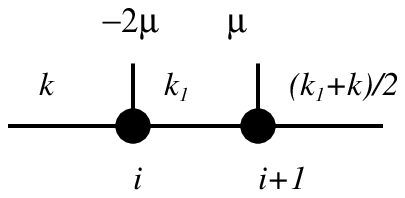}}
\right\rangle
=
\left\vert\vphantom{\frac{1}{1}}\right.\left.
\raisebox{-5mm}{\includegraphics[height=1.5cm]{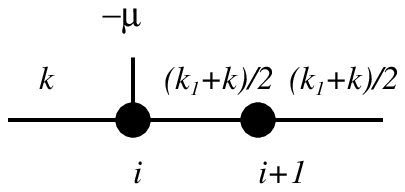}}
\right\rangle.\label{34}
\end{equation}
or the case in which a new vertex is created,
\begin{equation}
\hat{D}_i 
\left\vert\vphantom{\frac{1}{1}}\right.\left.
\raisebox{-5mm}{\includegraphics[height=1.5cm]{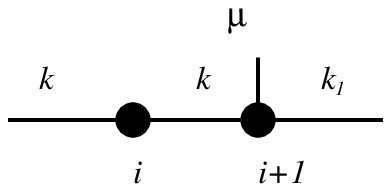}}
\right\rangle
=
\left\vert\vphantom{\frac{1}{1}}\right.\left.
\raisebox{-5mm}{\includegraphics[height=1.5cm]{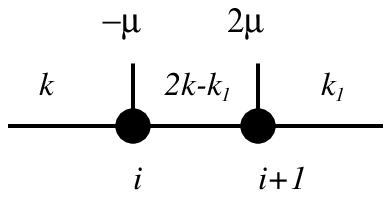}}
\right\rangle.
\end{equation}

To compute the adjoint of $\hat{D}$ is easy, since it is a one-to-one
operator. We start by noting that,
\begin{equation}
\left\langle
\raisebox{-5mm}{\includegraphics[height=1.5cm]{f3}}\right.
\left.\vphantom{\frac{1}{1}}\right\vert
\hat{D}_i \left\vert\vphantom{\frac{1}{1}}\right.\left.
\raisebox{-5mm}{\includegraphics[height=1.5cm]{f1}}
\right\rangle=1,
\end{equation}
and the insertion of any other bra in the left gives zero. 
Therefore
\begin{equation}
\hat{D}^\dagger_i 
\left\vert\vphantom{\frac{1}{1}}\right.\left.
\raisebox{-5mm}{\includegraphics[height=1.5cm]{f1}}
\right\rangle=
\left\vert\vphantom{\frac{1}{1}}\right.\left.
\raisebox{-5mm}{\includegraphics[height=1.5cm]{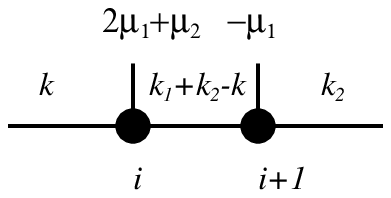}}
\right\rangle,
\end{equation}
with special particular cases that ``translate'' a $\mu$ insertion,
\begin{equation}
\hat{D}^\dagger_i 
\left\vert\vphantom{\frac{1}{1}}\right.\left.
\raisebox{-5mm}{\includegraphics[height=1.5cm]{f5}}
\right\rangle=
\left\vert\vphantom{\frac{1}{1}}\right.\left.
\raisebox{-5mm}{\includegraphics[height=1.5cm]{f4}}
\right\rangle,
\end{equation}
or create a vertex,
\begin{equation}
\hat{D}^\dagger_i 
\left\vert\vphantom{\frac{1}{1}}\right.\left.
\raisebox{-5mm}{\includegraphics[height=1.5cm]{f4}}
\right\rangle=
\left\vert\vphantom{\frac{1}{1}}\right.\left.
\raisebox{-5mm}{\includegraphics[height=1.5cm]{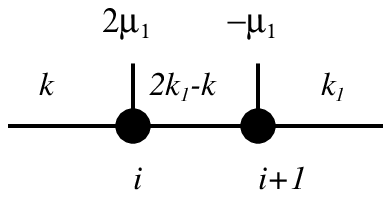}}
\right\rangle.
\end{equation}
In addition there is a third particular case of interest 
in which a vertex is annihilated,
it happens if $\mu_{-2}= -2\mu_1$ and $k=(k_1+k_2)/2$.

We now need to turn our attention to the other terms in the 
construction of $\hat{\mathbb H}$ in order to have a complete quantum
version of (\ref{27}). The discretization we  will propose is,
as,
\begin{equation}\label{discretizacion}
{\mathbb H} = \sum_{j=0}^N 
\left(O_{j+1}D_j-O_j\right)^\dagger
\left(O_{j+1}D_j-O_j\right)
\end{equation}
where $O_j =\sqrt[4]{\epsilon E^x_j}/(E^\varphi_j)^{3/2}$, and we have
chosen to localize $O_j$ and $D_j$ at different points. Intuitively
this can be seen in the fact that $\hat{D}$ ``shifts'' links in the
spin nets to the next neighbor whereas $\hat{O}$ just acts as a
prefactor, as we will discuss in the next paragraph. Therefore if one
wishes to find cancellations between both terms in
(\ref{discretizacion}) one needs to delocalize the action of both
$\hat{O}$'s.

The quantization of $O_j$ has been studied in the literature before
\cite{aspasi}. Since these operators only act multiplicatively, it
is better to revert to a simpler notation for the states
$\vert\vec{\mu},\vec{k}\rangle>$. The action of the operator is,
\begin{equation}
\frac{\sqrt[4]{\hat{E}^x_j}}{\left(E^\varphi_j\right)^{3/2}\epsilon^{1/4}}  
|\vec{\mu},\vec{k}>
= \left(\frac{4}{3\rho } \right)^6
\sqrt[4]{\frac{k_{j-1}+k_{j+1}}{2}} \left[\vert \mu_j+\frac{\rho}{2}\vert^{3/4} -
\vert \mu_j-\frac{\rho}{2}\vert^{3/4} \right]^6
|\vec{\mu},\vec{k}>,
\end{equation}
where $\rho$ is the minimum allowable value of $\mu$ as is customary in 
loop quantum cosmology. 
Since this operator has a simple action through a prefactor, we 
will call such prefactor $f(\vec{\mu},\vec{k},j)$.  One therefore has, for
example,
\begin{equation}
\hat{O}_{i+1} \hat{D}_i 
\left\vert\vphantom{\frac{1}{1}}\right.\left.
\raisebox{-5mm}{\includegraphics[height=1.5cm]{f4}}
\right\rangle=f(\vec{\mu},\vec{k},i+1)
\left\vert\vphantom{\frac{1}{1}}\right.\left.
\raisebox{-5mm}{\includegraphics[height=1.5cm]{f5}}
\right\rangle,
\end{equation}
or,
\begin{equation}
\hat{O}_{i+1} \hat{D}_i 
\left\vert\vphantom{\frac{1}{1}}\right.\left.
\raisebox{-5mm}{\includegraphics[height=1.5cm]{f1}}
\right\rangle=f(\vec{\mu},\vec{k},i+1)
\left\vert\vphantom{\frac{1}{1}}\right.\left.
\raisebox{-5mm}{\includegraphics[height=1.5cm]{f3}}
\right\rangle,
\end{equation}
where the $\vec{\mu},\vec{k}$ that appear in the prefactor are the ones
that appear in the state to the right of the prefactor.

It is worthwhile noticing that if $\mu_2=0$ the map is from a diagram with 
one insertion to another with one insertion, if $\mu_1=0$ it goes from
one insertion to two and if both $\mu_1$ and $\mu_2$ are non-vanishing
it maps two insertions to two insertions. It is not possible to go from
a state with two consecutive insertions into one with only one insertion,
since if $2\mu_2+\mu_1 =0$ then $f=0$. This is a key property one seeks in 
the regularization. If the regularization were able to fuse two insertions it
would be problematic, as we will discuss later on.

This allows us to evaluate the action of the quadratic Hamiltonian
${\mathbb H}$ explicitly on a set of states that capture in the
discrete theory the flavor of diffeomorphism invariance. For instance,
consider a normalized state obtained by superposing all possible
states with a given insertion
\begin{equation}
\left\vert \psi_1\right\rangle
=\frac{1}{\sqrt{N}}\sum_{i=0}^N 
\left\vert\vphantom{\frac{1}{1}}\right.\left.
\raisebox{-5mm}{\includegraphics[height=1.5cm]{f4}}
\right\rangle.
\end{equation}
Such a state would be the analogue in the discrete theory of a 
``group averaged'' state.
If we now consider the action of $\hat{O}_{i+1} \hat{D}_i-\hat{O}_i$ 
on such a
state we get,
\begin{equation}
\left\langle \psi_1\vphantom{\frac{1}{1}}\right\vert 
\hat{O}_{i+1} \hat{D}_i-\hat{O}_i 
\left\vert\vphantom{\frac{1}{1}}\right.\left.
\raisebox{-5mm}{\includegraphics[height=1.5cm]{f4}}
\right\rangle=0
\end{equation}
since both terms in the difference produce the same prefactor
when acting on the state on the right. If one were to consider on the
right a state with multiple insertions, then the result will also be 
zero since the operators do not convert two
consecutive insertions at $i,i+1$ into one
and the inner product would vanish. As a consequence,
we therefore have that,
\begin{equation}
\left\langle \psi_1\right\vert 
\hat{O}_{i+1} \hat{D}_i-\hat{O}_i 
=0.
\end{equation}

Let us now consider states with two insertions, again ``group averaged'' in the sense
that we sum over all possible locations of the two insertions respecting a relative
order within the lattice (in this case this is irrelevant due to cyclicity in a compact
manifold),
\begin{equation}
\left\vert \psi_2\right\rangle
=\frac{1}{\sqrt{N(N-1)}}\sum_{i=0}^N\sum_{\scriptstyle\begin{array}{c}\scriptstyle j
\neq i\\
\scriptstyle j=0\end{array}}^N 
\left\vert\vphantom{\frac{1}{1}}\right.\left.
\raisebox{-5mm}{\includegraphics[height=1.5cm]{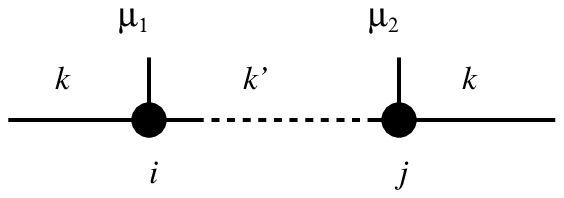}}.
\right\rangle.
\end{equation}

If one considers a state $\vert \nu \rangle$, with three or more 
insertions of $\mu$ one has that 
\begin{equation}
\left\langle \psi_2\right\vert 
\hat{O}_{i+1} \hat{D}_i-\hat{O}_i 
\left\vert \nu \right\rangle
=0,
\end{equation}
since in the first term $\hat{D}_i$ could produce a two insertion
diagram, but then the action of $\hat{O}$ at site $i+1$ would vanish,
and the term on the right does not produce a two insertion diagram,
as seen in (\ref{34}).
If one considers a state $\vert\nu\rangle$ with two non-consecutive
vertices, the operator also vanishes, for the same reasons as 
before. Finally, if $\vert\nu\rangle$ has two consecutive insertions
then we will have a non-trivial contribution. We will see, however,
that such a contribution vanishes in the continuum limit. To see this
we evaluate,
\begin{eqnarray}
\left\langle \psi_2\left\vert 
\hat{O}_{i+1} \hat{D}_i-\hat{O}_i 
\right\vert\vphantom{\frac{1}{1}}
\raisebox{-5mm}{\includegraphics[height=1.5cm]{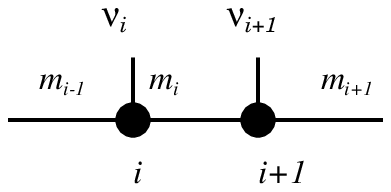}}
\right\rangle
&=&
f(\vec{\nu},\vec{m},i+1)
\left\langle \psi_2
\left\vert\vphantom{\frac{1}{1}}\right.\left.
\raisebox{-5mm}{\includegraphics[height=1.5cm]{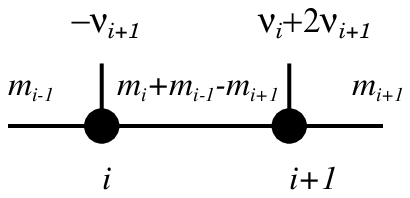}}\right.
\right\rangle\nonumber\\
&&
-f(\vec{\nu},\vec{m},i)
\left\langle \psi_2
\left\vert\vphantom{\frac{1}{1}}\right.\left.
\raisebox{-5mm}{\includegraphics[height=1.5cm]{f14}}\right.
\right\rangle\nonumber\\
&=&\left[
f(\vec{\nu},\vec{m},i+1)
\delta_{\mu_2,2\nu_{i+1}+\nu_i}
\delta_{\mu_1,-\nu_{i+1}}
\delta_{{k'},m_{i-1}}\delta_{{k'},m_i+m_{i-1}-m_{i+1}}
\delta_{k,m_{i+1}}\right.\nonumber\\
&&\left.-
f(\vec{\nu},\vec{m},i)
\delta_{\mu_1,\nu_i}
\delta_{\mu_2,\nu_{i+1}}
\delta_{k,m_{i-1}}
\delta_{{k'},m_{i}}
\delta_{k,m_{i+1}}\right]\frac{1}{\sqrt{N(N-1)}}
\end{eqnarray}

If $\vert \nu\rangle$ has one $\mu$ insertion then there is another
contribution,
\begin{eqnarray}
\left\langle \psi_2\left\vert 
\hat{O}_{i+1} \hat{D}_i-\hat{O}_i 
\right\vert\vphantom{\frac{1}{1}}
\raisebox{-5mm}{\includegraphics[height=1.5cm]{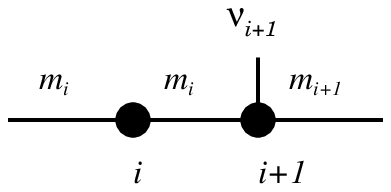}}
\right\rangle
&=&
f(\vec{\nu},\vec{m},i+1)
\left\langle \psi_2
\left\vert\vphantom{\frac{1}{1}}\right.\left.
\raisebox{-5mm}{\includegraphics[height=1.5cm]{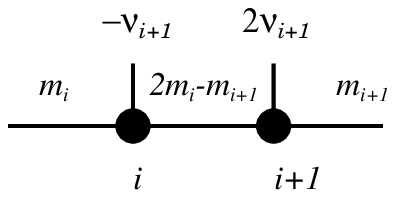}}\right.
\right\rangle\nonumber\\
&=&\frac{1}{\sqrt{N(N-1)}}\left[
\delta_{k,m_i}
\delta_{{k'},2m_i-m_{i+1}}
\delta_{\mu_1,-\nu_{i+1}}
\delta_{k,m_{i+1}}
f(\vec{\nu},\vec{m},i+1)\right]
\end{eqnarray}

We are now in a position to evaluate the expectation value of $\hat{\mathbb H}$. To do that we compute,
\begin{equation}
\langle \psi_2\vert\hat{\mathbb H} \vert \psi_2\rangle=
\sum_{j=0}^N 
\langle \psi_2\vert
\left(\hat{O}_{j+1} \hat{D}_j-\hat{O}_j\right)
\left(\hat{O}_{j+1} \hat{D}_j-\hat{O}_j\right)^\dagger
\vert \psi_2\rangle.
\end{equation}
and we insert a complete basis of states between 
the two parentheses. Then we can apply all the results we have 
just worked out. The final result is that only three finite contributions
appear for every $j$ and therefore
\begin{equation}
\langle \psi_2\vert\hat{\mathbb H}\vert \psi_2\rangle=O\left(\frac{1}{N}\right),
\end{equation}
and we see that in the limit $N\to \infty$ one shows that the spectrum of 
$\hat{\mathbb H}$ contains zero and therefore no anomalies appear and
the constraints are enforced exactly.

Analogously, one can show that for spin networks with $m$ vertices
$\langle\psi_m\vert{\mathbb H}\vert\psi_m\rangle=O(1/N)$, and
therefore the states that minimize $\langle\hat{\mathbb H}\rangle$ 
include in the limit
$N\to\infty$ the diffeomorphism invariant states obtained via the 
group averaging procedure.  To see this more clearly we note that the
state with $m$ vertices we are considering is of the form,
\begin{eqnarray}
\left\vert \psi_m\right\rangle =
\frac{1}{\sqrt{NC^N_m}}\sum_{i_{v_1}<\ldots<i_{v_j}<\ldots<i_{v_m}<i_{v_1}}
\left\vert\vphantom{\frac{1}{1}}
\raisebox{-5mm}{\includegraphics[height=1.5cm]{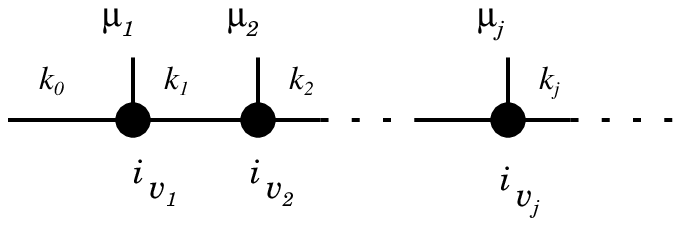}}
\right\rangle
\end{eqnarray}
where the sum is over all the spin nets with the only condition that
the cyclic order of the vertices is preserved, that is $v_1$ is
always between $v_m$ and $v_2$, etc. The quantities $C^{N-1}_m$ are
the combinatorial numbers of $N-1$ elements taken in groups of $m$ for
normalization purposes.
This sum is the discrete version of the sum on the group that is
performed in the continuum group averaging procedure.The sum preserves
the cyclic order placing the vertices in all the positions compatible
with such order.

We have shown that the expectation value of $\hat{\mathbb H}$ vanishes
in the continuum limit. Since $\hat{\mathbb H}$ is a positive definite 
operator this also implies that $\hat{\mathbb H}\vert \psi_n\rangle =0$, 
which is the condition one seeks in the uniform discretization approach.
This can be explicitly checked by computing, for instance for a 
state $\langle \psi_2\vert$,
\begin{equation}
\sum_s \langle \psi_2 \vert \hat{\mathbb H}\vert s\rangle\langle s\vert=
\frac{1}{\sqrt{N(N-1)}} \sum_{i=1}^N f_i \langle s_i\vert
\end{equation}
where the sum over $s$ means a sum over a basis of spin networks
$\vert s\rangle$ and the $\langle s_i\vert $  are spin network states
that have vertices at consecutive sites $i$ and $i+1$. Given that the
$f_i$'s are finite coefficients independent of $N$ one immediately sees
that the right hand side
has zero norm  when $N\to \infty$.

There is a rather important difference with the continuum case,
however. The states constructed here as limits of discrete states
are normalizable with the kinematical inner product and therefore
the calculation suggests that in a problem with a Hamiltonian
constraint in addition to diffeomorphism constraints one could work
all constraints in the discrete theory on an equal footing.

\section{Discussion}

We have seen in a $1+1$ dimensional model with diffeomorphism
invariance that one can discretize it, therefore breaking the
invariance, and treat it using the ``uniform discretizations'' approach
yielding a diffeomorphism invariant theory in the continuum limit.
We have argued that this would have been
close to impossible if one had naively discretized the constraints and
quantized the resulting theory.

An important point to realize is that the the kinematical Hilbert
space has been changed, by considering spin networks on ``lattices''
with a countable number of points. There exist infinitely many
possible such lattices built by considering different spacings between
the points.  However, in $1+1$ dimensions the choice of lattice does
not influence the diffeomorphism invariant quantum theory, whose
observables can be written in terms of the canonical variables and
invariant combinations of their derivatives that can be entirely
framed in terms of $\vec{k}$ and $\vec{\mu}$ without reference to
details of the lattice.  For instance, the total volume of a slice
evaluated on a diffeomorphism invariant spin network
$\vert\psi_1\rangle$ is given by
\begin{equation}
  \hat{V}\vert \psi_1 \rangle= 4 \pi  \ell_{\rm Planck}^3
\sum_v \vert \mu_v\vert \sqrt{\frac{k_{e^+(v)}+k_{e^-(v)}}{2}} \vert \psi_1 
\rangle
\end{equation}
where the sum is over all vertices of the continuum spin
network and $k_{e^\pm}$ are the values of $k$ emanating to the
right and left of vertex $v$. 

More generally, consider an observable $O_{\rm Diff}$, that is an
operator invariant under diffeomorphisms. Let us study in the space of 
lattices with a countable number of points its expectation
value on diffeomorphism invariant states $\langle \psi_{m,\vec{k},\vec{\mu}}
\vert \hat{O}_{\rm  Diff}\vert \psi_{m,\vec{k},\vec{\mu}}\rangle$, with 
$\vert \psi_{m,\vec{k},\vec{\mu}}\rangle$ the cyclic state we 
considered in the previous section. In the continuum the 
vectors of the Hilbert space of diffeomorphism invariant states 
$\vert \{s\}\rangle$ where $\{s\}$ is the knot class of a spin
network $s$ belong to the dual of the space of kinematic spin network states
$\vert s\rangle$. The expectation value of the observable in the
continuum is $\langle \{s\}\vert \hat{O}_{\rm  Diff} \vert \{s\}\rangle$
and the result of both expectation values in the continuum
and in the discrete theory coincide. The reason
for this is that the action of $\hat{O}_{\rm  Diff}$ on one of the 
terms in $\vert \psi_m \rangle$ coincides with $\hat{O}_{\rm Diff} 
\vert s\rangle$ except when $s$ has vertices that occupy consecutive
positions on the lattice. In this case, depending on the specific
form of $\hat{O}_{\rm Diff}$ the results could differ. Due to the
normalization factor, however, such exceptional contributions 
contribute a factor $1/N$ in the $N\to \infty$ limit, so we have
that in the continuum limit the expectation values in the continuum
and the discrete always agree. 

An issue of importance in loop quantum gravity is the problem of
ambiguities in the definition of the quantum theory. Apart from the
usual factor ordering ambiguities in a discrete theory one adds the
ambiguities of the discretization process. In this example we have
made several careful choices in this process to ensure that the
operator $\hat{\mathbb H}$ has a non-trivial kernel in the continuum
limit. This requirement proved in practice quite onerous to satisfy
and it took quite a bit of effort to satisfy the requirement. Though
in no way we claim that the results are unique, it hints at the fact
that requiring that $\hat{\mathbb H}$ have a non-trivial kernel in the
continuum significantly reduces the level of ambiguities in the
definition of a quantum discrete theory. We have not been able 
to find another regularization satisfying the requirement an leading
to a different non-trivial kernel.

Another point to note is that the quantum diffeomorphism constraints
$\phi^\epsilon(M)=\sum_{j=0}^N
\frac{M_j}{2i\epsilon}\left(D_j-1\right)$ with $M_j$ stemming from
discretizing a smooth shift function do not reproduce the continuum
algebra of constraints when they act on generic spin networks on the
lattice that belong to the kinematical Hilbert space. The algebra
almost works, but there appear anomalous contributions for spin
networks with vertices in two consecutive sites of the lattice.  In
spite of this the constraints can be imposed at a quantum level
through the condition $\langle \psi \vert H=0$ and imply, as we
showed, that the solutions correspond to a discrete version of the sum
in the group that is performed in the group averaging procedure.  The
difference is that these states are normalizable with the inner
product of the kinematical space itself. In this construction the
Hilbert space ${\cal H}_{\rm Diff}$ is a subspace of ${\cal H}_{\rm
Kin}$, unlike the situation in the ordinary group averaging procedure.
This property opens interesting possibilities, particularly if it
holds in more elaborate models. If such a property were to hold in
more complex models, for instance involving a Hamiltonian constraint,
it would be very important since it would provide immediate access
to a physical inner product.

All of the above suggests that in more realistic models than the one
we studied, for instance when there is a Hamiltonian constraint (with
structure functions in the constraint algebra) one will also be able
to define the diffeomorphism and the Hamiltonian constraints as
quantum operators and impose them as constraints (or equivalently, to
impose the ``master constraint'' ${\mathbb H}$). They would act on the
kinematic Hilbert space of the discrete theory, and one would hope
that a suitable continuum limit can be defined.  We would therefore
have a way of defining a continuum quantum theory via discretization
and taking the continuum limit even in systems where the
discretization changes the nature of the constraints from first to
second class. In $1+1$ dimensions the procedure appears quite
promising. It should be noted that this is a quite rich arena in
physical phenomena, including Gowdy cosmologies, the Choptuik
phenomena and several models of black hole formation. The fact that we
could envision treating these problems in detail in the quantum theory
in the near future is quite attractive. In higher dimensions the
viability of the approach will require further study, in particular
since the discretization scheme chosen could constrain importantly the
types of spin networks that one can construct in the continuum theory.

Summarizing, we have presented the first example of a model with
infinitely many degrees of freedom where the uniform discretization
procedure works out to the last consequences, providing a continuum
theory with diffeomorphism invariance and where the master constraint
has a non-trivial kernel. It also leads to an explicit construction of the
physical Hilbert space that is different from the usual one, allowing the
introduction of the kinematical inner product as the physical one.

\section{Acknowledgements}

This work was supported in part by grants NSF-PHY0650715, and by funds
of the Horace C. Hearne Jr. Institute for Theoretical Physics, FQXi,
PEDECIBA and PDT \#63/076 (Uruguay) and CCT-LSU.

\end{document}